\documentclass[aps,prb,twocolumn,showpacs]{revtex4}
\usepackage{graphicx}

\begin{document}
\title{Electronic and optical properties of \mbox{LiBC}}
\author{A. V. Pronin$^{1,2,}\footnote {Electronic address: artem.pronin@physik.uni-augsburg.de}$,
K. Pucher$^{1}$, P. Lunkenheimer$^{1}$, A. Krimmel$^{1}$, and A.
Loidl$^{1}$}
\address{$^{1}$Experimentalphysik V, Elektronische Korrelationen und Magnetismus,
Institut f\"{u}r Physik, Universit\"{a}t Augsburg, 86135 Augsburg, Germany \\
$^{2}$Institute of General Physics, Russian Academy of Sciences,
119991 Moscow, Russia}
\date{\today}

\begin{abstract}

LiBC, a semiconducting ternary borocarbide constituted of the
lightest elements only, has been synthesized and characterized by
x-ray powder diffraction, dielectric spectroscopy, and
conductivity measurements. Utilizing an infrared microscope the
phonon spectrum has been investigated in single crystals. The
in-plane B-C stretching mode has been detected at 150 meV,
noticeably higher than in AlB$_{2}$, a non-superconducting
isostructural analog of MgB$_{2}$. It is this stretching mode,
which reveals a strong electron-phonon coupling in MgB$_{2}$,
driving it into a superconducting state below 40 K, and is
believed to mediate predicted high-temperature superconductivity
in hole-doped LiBC [H. Rosner, A. Kitaigorodsky, and W. E.
Pickett, Phys. Rev. Lett. \textbf{88}, 127001 (2002)].
\end{abstract}


\maketitle

The discovery of phonon-mediated superconductivity in MgB$_2$ with
a superconducting transition temperature of $T_c \sim 40$ K (Ref.
\onlinecite{nagamatsu}) has triggered enormous interest in similar
layered compounds. MgB$_2$ consists of graphite-like boron layers
with the Mg ions intercalated in between. The B-B bonding
$\sigma$-bands reveal a two dimensional character and are strongly
coupled to the in-plane stretching mode, consistent with a
phonon-mediated BCS type pairing mechanism.\cite{an} In Raman and
neutron scattering experiments \cite{bohnen, renker} an extremely
strong reorganization of this stretching mode has been
unambiguously demonstrated. The authors found that the in-plane
boron vibration in MgB$_{2}$ is shifted to 70 meV, while in the
isostructural but non-superconducting AlB$_2$ it is located ar 123
meV. From the averaged phonon density and the electron-phonon
coupling the superconducting transition temperatures can be
estimated and the theoretical estimates using a strong
electron-phonon coupling deduced superconducting transition
temperatures of the correct magnitude.\cite{bohnen, kong} Similar
layered compounds have been theoretically investigated
\cite{ravindran, rosner, singh} and a superconducting transition
temperature of $T_c \sim 120$ K has been predicted for hole-doped
LiBC by Rosner \textit{et al}.\cite{rosner}

Since the mid 80's LiBC was a subject of only a few experimental
studies.\cite{mair, woerle} It reveals the same structure as
MgB$_2$, but with the planar hexagonal boron layers replaced by
B-C layers, with the B and C atoms alternating both within the
layer and along the \textit{c} direction. LiBC has $P6_3/mmc$
symmetry with $a=0.275$ nm and $c=0.7058$ nm.\cite{woerle} The
stoichiometric compound is insulating. From band structure
calculations a gap of at least 1~eV has been deduced for the pure
compound and it has been shown that on hole doping, induced by
Li-deficiency, the $\sigma$ bands significantly contribute to the
Fermi level and are strongly coupled to the B-C stretching
mode.\cite{rosner}

In this work we synthesized powder and single crystalline LiBC,
characterized the samples by x-ray powder diffraction, determined
the dielectric constant and the conductivity as function of
temperature, and measured the optical properties in the
far-infrared and mid-infrared range. Besides the interesting
question of the phonon renormalization in hole-doped compounds,
which should be the subject of further studies, it certainly also
is highly interesting to investigate a ternary compound
synthesized from the lightest elements only.

\begin{figure}[t]
\includegraphics[width=\columnwidth,clip]{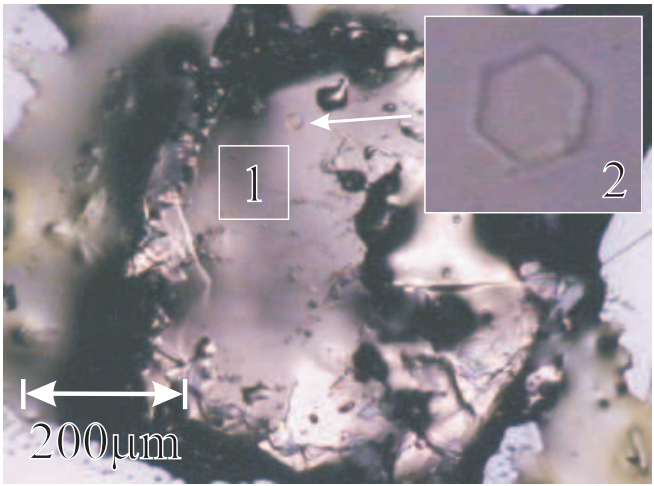}
\caption{Photograph of a LiBC crystalline sample. (1) shows an
example of the area investigated in the infrared experiments. The
inset (2) shows a magnification of a little terrace on the sample
surface with a perfect hexagonal shape.} \label{fig1}
\end{figure}

LiBC was synthesized from the elements at 770 K with subsequent
annealing at 1770 K in sealed niobium ampoules with open
molybdenum crucibles inside. The molybdenum crucibles, 5 mm
diameter and approximately 50 mm length, have been filled in an
argon box with commercially available lithium, boron, and carbon
with a lithium excess of 300 \% compared to the stoichiometric
ratio. This excess is necessary to provide a high lithium vapor
pressure in the ampoule. The carbon and boron powders were
preliminarily annealed in vacuum.  The Mo crucibles have been
placed in Nb ampoules, the latter sealed by argon welding. It is
known, that molybdenum does not react with any of the components,
while it can not hold high pressures at high temperatures. Due to
this fact an additional niobium ampoule is necessary. The use of a
Nb ampoule only is impossible since niobium reacts with boron.
This sample preparation technique has been developed in the groups
of R. Nesper (Z\"{u}rich) and H. G. von Schnering (Stuttgart), and
is described in detail in Refs. \onlinecite{mair, woerle}.  The
ampoules have been heated up to 770 K with a rate 100 K/hour, held
at 770 K during one hour, then heated up to 1770 K with 200
K/hour, held one more hour at this temperature, and finally cooled
down with a rate of approximately 150 K/hour. Faster cooling seems
to support the growth of crystals of larger size. After removing
the extra lithium the product consists of mainly yellow golden
powder. Some crystals can be quite large reaching the size of $0.5
\times 0.5 \times 0.01$~mm$^3$. One of these platelets is shown in
Fig. \ref{fig1}. The hexagonal morphology is clearly indicated by
the terraced ideal hexagon shown in the inset 2 in Fig.
\ref{fig1}.

The x-ray powder diffraction measurements have been performed at
room temperature using a STOE diffractometer with a position
sensitive detector utilizing Cu-K$_\alpha$ radiation with
$\lambda=0.1541$ nm. The data have been analyzed by standard
Rietfeld refinement. The diffraction pattern has shown the correct
symmetry ($P6_3/mmc$) with no indications of spurious phases. The
lattice constants, $a=0.2751(1)$~nm and $c=0.7055(1)$~nm, were
found in good agreement with published results.\cite{woerle}

\begin{figure}[t]
\centering
\includegraphics[width=\columnwidth,clip]{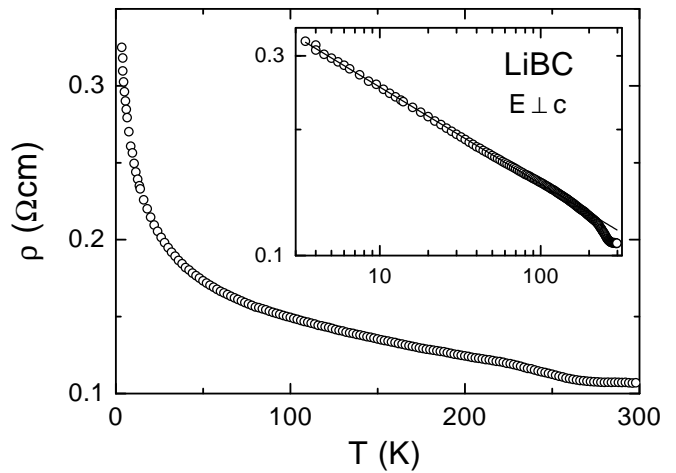}
\caption{Temperature dependence of the dc resistivity of LiBC,
measured with the field parallel to the BC planes using four-point
contact configuration. The inset shows the same data in a
double-logarithmic representation, the line demonstrating an
approximate $T^{-n}$ power-law behavior with $n \approx 0.23$.}
\label{fig2}
\end{figure}

The dc-resistivity of a single-crystalline sample was measured
using a standard four-point technique in the range 3 K $< T <$ 300
K. Using a coplanar contact geometry the electrical field was
directed parallel to the B-C layers. The contacts were prepared by
silver paint. Due to the ill-defined geometry of the rather small
sample, the absolute values of the resistivity should be taken as
a rough estimate only. As is shown in Fig. \ref{fig2}, the
temperature dependence of the resistivity exhibits semiconducting
characteristics, however, only with a rather small overall change
of about a factor of three in the investigated temperature range.
It does not follow a thermally activated behavior, $\rho \propto
exp(E/T)$. Also the prediction of the variable range hopping
model, $\rho \propto exp(B/T^{1/4})$, which often can be applied
for hopping conduction of localized charge carriers \cite{mott},
does not work for LiBC. Instead, in a rather broad temperature
range the resistivity can be phenomenologically described by a
power law, $\rho \propto T^{-n}$, as demonstrated in the
double-logarithmic representation in the inset of Fig. \ref{fig2}
revealing an exponent $n \approx 0.23$. However, we are not aware
of any theoretical model that could explain such a power law
behavior.

\begin{figure}[t]
\centering
\includegraphics[width=7cm,clip]{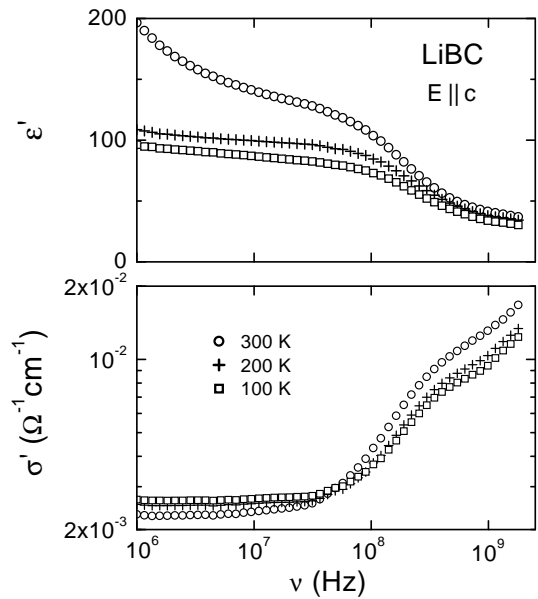}
\vspace{0.2cm} \caption{Frequency dependence of the real parts of
the dielectric constant $\varepsilon'$ and ac conductivity
$\sigma'$ of LiBC for three temperatures. The measurements were
performed using a two-point contact configuration with the
electric field directed parallel to the B-C planes.} \label{fig3}
\end{figure}

In addition, the dielectric properties of the platelet-shaped
samples were measured in sandwich geometry, i.e. with the field
directed perpendicular to the B-C layers. The measurements were
performed for frequencies $10^{6} \leq \nu \leq 2\times10^{9}$ Hz
employing a reflectometric technique, described in detail in Ref.
\onlinecite{BoehmerSchneider}. Fig. \ref{fig3} shows the frequency
dependence of the dielectric constant $\varepsilon'$ and the
conductivity $\sigma'$ for three temperatures. $\varepsilon'(\nu)$
exhibits a step-like decrease with the point of inflection close
to 300 MHz, $\sigma'(\nu)$ an increase with a shoulder near the
same frequency. Such a behavior could be indicative of a dipolar
relaxation process, but as the existence of dipolar entities in
LiBC is unlikely, a non-intrinsic origin of the observed
relaxation-like features seems to be more reasonable. Thus we
ascribe this behavior to Maxwell-Wagner type contributions of the
interface between sample and contacts. As demonstrated in detail
in Ref. \onlinecite{Lunki}, then the intrinsic response is
revealed at high frequencies only, when the contact resistance is
bridged by the contact capacitance acting like a short at high
frequencies. From Fig. \ref{fig3}, close to 1 GHz we read off a
value of $\varepsilon' \approx 35$, which smoothly saturates
towards higher frequencies and is nearly temperature independent.
From the frequency dependence of $\sigma'(\nu)$ in the intrinsic
region at $\nu > 300$ MHz a plateau, followed by a further
increase, can be suspected, which would be indicative of hopping
conduction of localized charge carriers.\cite{Lunki} Measurements
at even higher frequencies are necessary to corroborate this
conjecture. Within this picture the plateau represents the
intrinsic dc conductivity, which can be read of as $\sigma_{dc}
\approx 10^{-2}$ $\Omega^{-1}$cm$^{-1}$, showing a weak
temperature dependence only. Thus the dc resistivity perpendicular
to the BC planes seems to be more than two orders of magnitude
higher than the measured dc resistivity parallel to the planes
(cf. Fig. \ref{fig2}) indicating a strong anisotropy of the
electrical transport properties.

The measurements of the optical reflectivity were carried out
using the infrared microscope Bruker IRscope II, which was
connected to a Fourier-transform infrared spectrometer Bruker IFS
66v/S. The measurements were carried out on flat surfaces of
as-grown single crystals at frequencies from 400 to
2000~cm$^{-1}$. In order to eliminate multiple reflections from
the opposite sides of the plane-parallel samples, the samples were
tilted by a small angle (approximately 10$^{\circ}$) to the normal
incidence. Using the infrared microscope we were able to focus on
a small fraction of the LiBC platelets with the best flatness
compared to surrounding areas. An example of such an investigated
area is shown in Fig. \ref{fig1}.

\begin{figure}[t]
\centering
\includegraphics[width=\columnwidth,clip]{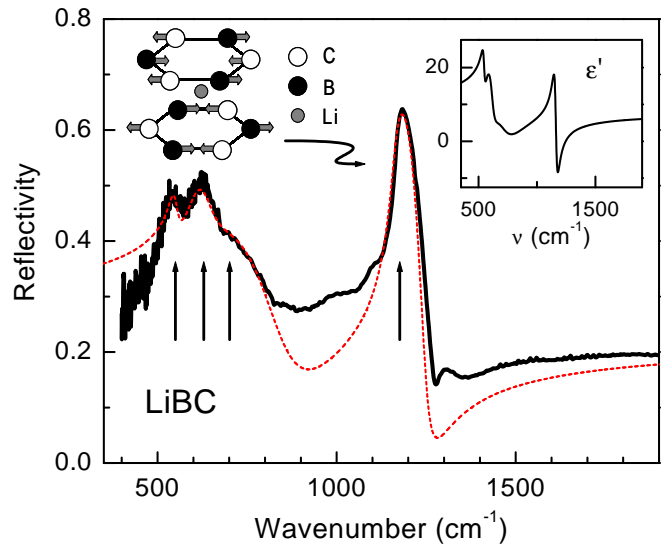}
\caption{Reflectivity of a LiBC crystal in the far- and
mid-infrared range. The vertical arrows mark the phonon positions.
The dashed line is a fit of the experimental  spectra with four
Lorentz oscillators. The inset shows the dielectric permittivity
calculated from the fit of the reflectivity data. The displacement
pattern of the B-C stretching infrared-active mode is shown in the
upper left corner.} \label{fig4}
\end{figure}

The reflectivity as obtained from the infrared measurements is
shown in Fig. \ref{fig4}. We observe three well-defined peaks in
the spectrum at 540, 620, and 1180 cm$^{-1}$, and assign them to
the infrared active phonons. A large shoulder at 700 cm$^{-1}$
most likely represents the forth optical phonon. It should be
noted that the quasiperiodic variation of $R$, visible in the
spectra, is the residual influence of the multiple reflections
inside the sample, which could not be completely removed by
tilting the sample. This interference pattern is most clearly seen
at 1300 cm$^{-1}$, and also between 800 and 1100 cm$^{-1}$.
Spectra measured closer to the normal incidence reveal much more
pronounced fringes. These oscillations make a quantitative
analysis quite difficult, and the fit with four Lorenzians shown
in Fig. \ref{fig4} by a dashed line is not ideal. The inset in
Fig. \ref{fig4} shows the dielectric permittivity $\varepsilon'$
as calculated from the fit of the reflectivity spectrum. Since the
description of the reflectivity is not perfect, the data for
$\varepsilon'$ are subject to large errors. However, we would like
to stress that $\varepsilon' \approx 15$ in the terahertz regime
is not too far from $\varepsilon' \approx 35$ as measured at 1
GHz, having in mind that the intrinsic value still is not quite
reached at this frequency. The remaining difference most likely is
due to a relative large uncertainty in the determination of the
crystal thickness, which affects the absolute values of
$\varepsilon'$ in the dielectric measurements.

The group analysis for LiBC ($P6_3/mmc$, $D^{4}_{6h}$) yields the
following modes at the $\Gamma$-point: $3A_{2u} + 2B_{1g} + B_{2u}
+ 3E_{1u} + E_{2u} + 2E_{2g}$, four of which are optical
infrared-active modes ($2A_{2u}$ and $2E_{1u}$). The four modes
observed in our measurements most likely represent these
infrared-active phonons. Although the reflectance was measured
from quite plane sample surfaces, which are supposed to be mostly
parallel to the B-C planes, we believe that a significant amount
of the out-of-plane response is also present in our measurements.
The tilted geometry further increases the out-of-plane
contribution.

If one compares the observed absolute values of the phonon
frequencies in LiBC with those observed and calculated for
aluminum diboride \cite{bohnen, renker}, one immediately finds a
close similarity. In both compounds there is a very high frequency
phonon, which is a Raman-active $E_{2g}$ mode in AlB$_{2}$ at
$\sim$ 980 cm$^{-1}$ (123 meV) and the mode at 1180 cm$^{-1}$
($\sim$ 150 meV) in our measurements of LiBC. We ascribe this mode
to the $E_{1u}$ infrared-active phonon. Due to the doubling of the
unit cell in LiBC compared to AlB$_{2}$, the stretching
Raman-active $E_{2g}$ mode of the B-B hexagons transforms into two
modes in LiBC: $E_{2g}$ and $E_{1u}$. The latter is
infrared-active, its displacement pattern is shown in Fig.
\ref{fig4}. The $E_{2g}$ mode in AlB$_2$ has been shown to soften
considerably on doping this compound with Mg, which is accompanied
by the onset of superconductivity in Mg$_{1-x}$Al$_{x}$B$_2$ (Ref.
\onlinecite{renker}). This softening is believed to result from
the electron-phonon renormalization driving the superconductivity
in magnesium diboride.

The three other infrared modes in LiBC have lower frequencies and
are situated quite close to each other, resembling again the
situation in AlB$_{2}$, where two infrared-active modes are
supposed to exist in the range of 300 to 500 cm$^{-1}$ (Refs.
\onlinecite{bohnen, renker}). The moderate upward shift in the
frequencies of the observed modes in LiBC compared to the modes in
AlB$_{2}$ is due to the fact that LiBC consists of lighter
elements. Concluding this section, we mention that we have
extended the infrared measurements up to approximately 1 eV and
found no indication of the onset of an interband transition,
demonstrating that the band gap is substantially larger than 1 eV.
In the LDA calculations of Rosner \textit{et al}. valence and
conduction bands are separated by 1 eV. By W\"{o}rle \textit{et
al}. \cite{woerle} band gaps between 2.5 and 4.2 eV have been
calculated.

In conclusion, we have synthesized the ternary borocarbide LiBC,
characterized it by x-ray diffraction, and performed dc
resistivity and dielectric measurements on single crystals. The dc
resistivity demonstrates semiconducting temperature
characteristics with strong anisotropy of $\rho _{\parallel c} /
\rho _{\perp c} \geq 100$. Infrared spectroscopy on single
crystals utilizing an infrared microscope reveals four
infrared-active phonon modes, consistent with point-group symmetry
considerations. The high-frequency mode at 150 meV can be ascribed
to the B-C stretching mode. It is this phonon mode that is
expected to drive phonon-mediated superconductivity in hole-doped
LiBC at temperatures substantially higher than in MgB$_{2}$ (Ref.
\onlinecite{rosner}). The next step is to follow the softening of
this in-plane mode as a function of Li deficiency.

\begin{acknowledgments}

We would like to thank U. Killer, F. Mayr, M. M\"{u}ller, A.
Pimenova, and D. Vieweg for the experimental support. A. V. P. is
also grateful to J. Hlinka and J. Petzelt for valuable
consultations. This work was partly supported by the DFG via the
Sonderforschungsbereich 484 (Augsburg) and by the BMBF via the
Contract No. EKM/13N6917/0.

\end{acknowledgments}

\end{document}